\documentclass[twocolumn,showpacs,preprintnumbers,amsmath,amssymb]{revtex4}

\usepackage{graphicx}
\usepackage{dcolumn}
\usepackage{bm}

\begin{document}

\preprint{APS/123-QED}

\title{Formation of the internal structure of solids under severe action}

\author{Leonid S.Metlov}

 \email{lsmet@kinetic.ac.donetsk.ua}
\affiliation{Donetsk Institute of Physics and Engineering, Ukrainian
Academy of Sciences,
\\83114, R.Luxemburg str. 72, Donetsk, Ukraine
}%

\date{\today}

\begin{abstract}
On the example of a particular problem, the theory of vacancies, a new form of kinetic equations symmetrically incorporation the internal and free energies has been derived. The dynamical nature of irreversible phenomena at formation and motion of defects (dislocations) has been analyzed by a computer experiment. The obtained particular results are extended into a thermodynamic identity involving the law of conservation of energy at interaction with an environment (the 1st law of thermodynamics) and the law of energy transformation into internal degree of freedom (relaxation). The identity is compared with the analogous Jarzynski identity. The approach is illustrated by simulation of processes during severe plastic deformation, the Rybin kinetic equation for this case has been derived.
\end{abstract}

\maketitle

\section{Introduction}

Thermodynamics of internal state variables has a long history. An important idea to take into account internal microstructure changes for polyatomic gas by introducing an additional variable was proposed by Herzfeld and Rice in 1928 \cite{hr28}. Later, the concept was used by Mandelshtam and Leontovich for solving the same problem for liquids \cite{ml37}. Supplemented with elements of rational mechanics the concept took a completed and closed form in the works by Coleman and Gurtin \cite{cg67}. The concept was given an alternative development in the works by Landau. On the base of newly introduced internal state variable in a form of the order parameter a theory of phase transitions was constructed by him at the quality level \cite{lh54,ll69}.

Studies of the problem have proposed a large variety of kinetic equations describing changes in the internal structure under nonequilibrium conditions within the conceptual framework of constitutional theories \cite{za87,b05}, continuum theories of defects \cite{e56,k58,ce83,sw04}, extended continuum theories \cite{kkbt02,fs03}, self-organization theories of dislocations \cite{gll95,m99,vbf05},  gradient \cite{ga99,g06} and non-local theories \cite{bj02,pgbb02}, mesoscopic theories \cite{dklm01,rg03}, and so on. There is a special need to mention the works of Landau's School \cite{lh54,ll69}, which was developing one of the most physical lines of the theory of non-equilibrium phenomena. In the modern science the line is presented by phase-field theories \cite{akv00,esrc02,lpl02,rbkd04,gptwd05,akegny06,rjm09}. Unlike Landau theory, the order parameter of phase-field theory is not well-defined, but it does not hinder the obtaining results which are in a good accordance with the behavior of real systems.

The order parameter introduced by Landau is a type of the internal variable and therefore, it should describe changes in real structure state of a system. However, it describes the state in terms of structure itself not directly but through an immediate indicator connected with symmetry change. It may turn out that in some applications it is more convenient to directly connect a system state with the change of a specific structural parameter, for example, the density of defects.

Internal phenomena proceeding in a solid represent a complex pattern of mutual transformations of the energy from each type to other ones. In most cases the physical nature of internal phenomena is not defined concretely \cite{b64}. It is assumed that all of them are of fluctuation origin \cite{yk67,cw51}. At the same time, the generation and motion of structural defects in a solid under a severe external action can be also regarded as some internal processes characterized by a set of internal variables \cite{c05}. These channels of energy dissipation still differ from traditional viscous channels of dissipation as the relaxation involving the defects has the dynamical nature. It doesn't run directly by means of transition of the organized energy to chaotic thermal one but passes a series of intermediate levels and stages. During generation of defects  a part of organized (elastic) energy is spent directly for the formation of structural defects which, along with the thermal channel, make parallel channels of dissipation.

The other part of the energy transforms to thermal form not immediately, it is initially radiated in the form of low-frequency vibrations and waves (acoustic emission) which can be treated as a part of the general dissipation phenomenon \cite{ptz07,i04,k06}. Thereupon this part of the energy is scattered on the equilibrium thermal vibrations (waves) replenishing their energy. So, the energy stored in the non-equilibrium subsystem cannot be too much.

The present paper is devoted to the development of the Landau approach. In Sec. II on the example of the vacancy theory we propose a new form of kinetic equations for defects which is symmetrical with respect to free and internal energy utilization. In this section invoking molecular dynamics simulation, we select equilibrium and nonequilibrium heat subsystems of a solid under severe load and introduce a notion of nonequilibrium temperature. In Sec. III the results obtained in the previous sections are generalized in the form of a thermodynamic identity combining the first and second laws of thermodynamics for the described processes. In Sec. IV the general results are discussed. Section V is devoted to the application of kinetic equations to the description of grain refinement during severe plastic deformation. Our conclusions are presented in Section VI.

\section{KINETIC EQUATIONS}

To begin the analysis of the problem it is convenient to maximally simplify it. Let us split the problem into two parts, the first, structural one connected with generation of defects and the second, thermal part, and consider them separately. Moreover, we consider the structural part of the problem on the simplest example of a solid with vacancies.

\subsection{Structural part of the problem}

A classical theory of vacancies was developed by Frenkel and Kittel \cite{f55,k67}. In the variant proposed by Frenkel, the internal energy of a solid with vacancies is presented in the form $U=U_{0}+u_{V}n$, where $U_{0}$ is the internal energy of the vacancy-free crystal, $u_{V}$ is the average vacancy energy, $n$ - is the number of vacancies. Next, the transformation to the configurational free energy $F_{c}=U-TS_{c}$, where $T$ is the temperature, $S_{c}$ is the configurational entropy, is performed. The configurational entropy is uniquely expressed through the number of vacancies, and, at low vacancy density, it is true $S_{c}=kN(lnN/n+1)$, where $k$  is the Boltzmann’s constant, $N$  is the total number of lattice sites occupied by atoms. The equilibrium value of the number of vacancies $n_{e}$ is found from the free energy minimum provided the vacancy energy is independent of the number of vacancies, that is, $u_{V}=u_{V0}$, where $u_{V0}$ is a constant value for a given matter. The free energy minimization procedure leads to the equation of state $n_{e}=N exp(-(u_{V0}/kT))$.

\begin{figure*}
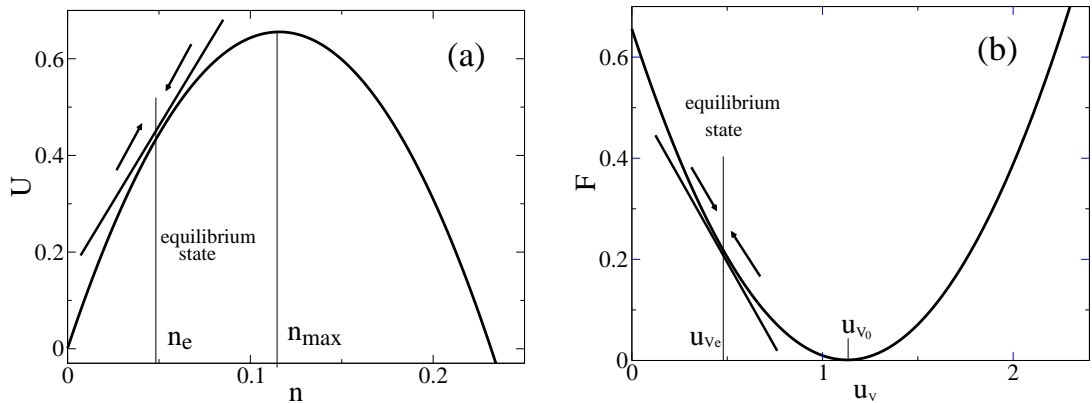

\includegraphics [width=2.7 in]{fig_1a}
\quad \quad
\includegraphics [width=2.7 in]{fig_1b}
\caption{\label{f1} Plots of the internal (a) and free (b) energy versus its eigen arguments. Tendency of the system to the equilibrium state is indicated by arrows.}
\end{figure*}

With the vacancy energy dependence on vacancy number taken into account we consider a quadratic correction to the internal energy in the form:
  \begin{equation}\label{a1}
U=U_{0}+u_{V0}n-\dfrac{1}{2}u_{V1}n^{2},
  \end{equation}
 where $u_{V1}$ is a coefficient. Since the energy needed for the formation of a new vacancy, in conditions when in a solid there are already a number of other vacancies, is less than in the case of the vacancy-free crystal, then the negative sign must be used for the correction term. Note that expression (1) is true both for equilibrium and non-equilibrium states. In this approximation the internal energy is a convex function of the defect number having the maximum at point $n=n_{max}$, as it is shown in Fig. \ref{f1}a.
 For the transition to the free energy with the Legendre's transformation, it is necessary to subtract the bound energy $TS_{c}$  from the internal one. According to Kittel, a theory of vacancies can be equivalently formulated either in terms of the configurational entropy or the chemical potential \cite{k67}. In this context, the bound energy can be written down in two equivalent forms:
  \begin{equation}\label{a2}
TS_{c}=u_{V}n.
  \end{equation}

Strictly speaking, relationship (\ref{a2}) is just only in the equilibrium form, as the equilibrium temperature $T$ is contained there. Following the Landau style, we introduce the generalized thermodynamic force for non-equilibrium state in the form:
  \begin{equation}\label{a3}
u_{V}=\dfrac{\partial U}{\partial n}=u_{V0}-u_{V1}n.
  \end{equation}

Relationship (\ref{a3}) can be treated as an equation of state true for both the equilibrium state and non-equilibrium cases. Applying, though formally, the Legendre transformation with respect to a pair of the conjugate variables $u_{V}$ and $n$ and with the equation of state (\ref{a3}) one obtains for the free energy:
  \begin{equation}\label{a4}
F_{c}=U_{0}+\dfrac{1}{2u_{V1}}(u_{V0}-u_{V})^{2}.
  \end{equation}

Note that an eigen argument of the internal energy is the vacancy number (density), and the eigen argument of the configurational free energy is the vacancy energy. In this approximation the free energy is a concave function with the minimum at point $u_{V}=u_{V0}$, as it is shown in Fig. \ref{f1}b.

We define the equilibrium state from a macrocanonical distribution taking it in the form:
  \begin{equation}\label{a5}
f(n)=C\dfrac{(N+n)!}{N!n!}\exp(-\dfrac{U(n)}{kT}),
  \end{equation}

The pre-exponential factor describes the combinational, that is, entropy, part of the distribution function connected with degeneration of microstates. The exponential factor describes the restrictive part of the distribution function connected with the overcoming of potential barrier between microstates. The most probable state determined from condition $\partial f(n)/\partial n=0$ gives the equation of state in the form:
  \begin{equation}\label{a6}
n_{e}=N\exp(-\dfrac{u_{V}}{kT}),
  \end{equation}
which is true only for the equilibrium case. As for the equilibrium state the equations of state (\ref{a6}) and (\ref{a3}) must be identical, then substituting Eq. (\ref{a6}) into (\ref{a3}) one gets a condition for the determination of this state:
  \begin{equation}\label{a7}
kT \ln\dfrac{N}{n_{e}}=u_{V0}+(\dfrac{kT}{2N}-u_{V1})n_{e},
  \end{equation}

With relationship (\ref{a7}), it is easy to show that neither the maximum of the internal energy nor the minimum of the free energy satisfy this state. The equilibrium state is at point $n=n_{e}$, where relationships 
  \begin{equation}\label{a8}
u_{Ve}=\dfrac{\partial U}{\partial n_{e}},  
\quad n_{e}=-\dfrac{\partial F_{c}}{\partial u_{Ve}},
  \end{equation}
are valid. Here the additional subscript $e$ denotes the equilibrium value of a variable.

If a system has deviated from the equilibrium state then it should tend to that state with a speed which is the higher, as the deviation is larger:
  \begin{equation}\label{a9}
\dfrac{\partial n}{\partial t}=\gamma_{n}(\dfrac{\partial U}{\partial n}-u_{Ve}),  
\quad \dfrac{\partial u_{V}}{\partial t}=-\gamma_{u}(\dfrac{\partial F_{c}}{\partial u_{V}}-n_{e}),
  \end{equation}

The both variants of the kinetic equations are equivalent and their application is dictated by convenience reasons. The form of kinetic equations (\ref{a9}) is symmetric with respect to the use of internal and configurational free energy. In the right-hand part of Eq. (\ref{a9}) the signs are chosen basing on solution stability, the internal energy is a convex function, and the free energy is a concave one. For the well-known Landau-Khalatnikov kinetic equation
  \begin{equation}\label{a10}
\dfrac{\partial n}{\partial t}=-\gamma \dfrac{\partial F_{c}}{\partial n},
  \end{equation}
the contradiction is typical. It is written in terms of the free energy, but the evolved variable   is not an eigen argument of the free energy. However, one could show that Eq. (\ref{a10}) can be derived from the second Eq. (\ref{a9}) in the limit of independence of the defect energy on the defect number $u_{V1}\rightarrow 0$.

Thus, the Landau-Khalatnikov equation can be only generalized to other types of defects with the above condition satisfied. In the general case, when the defect energy depends on the defect number and strongly changes in a studied process, for example, in the case of severe plastic deformation in metals with grain boundaries as defects, it is necessary to use a more general form of the kinetic equations (\ref{a9}). Peculiarities of the thermal part of the problem can be considered for a special case a dislocation generation during the indentation.

\subsection{Thermal part of the problem}

Time scans of the wave-like motion of lattice atoms can be examined as some signals. Every atom participates simultaneously in both the equilibrium and non-equilibrium motions. Using the differences in frequency properties they can be separated by means of an ordinary filtration.

\begin{figure}
\includegraphics [width=2.7 in]{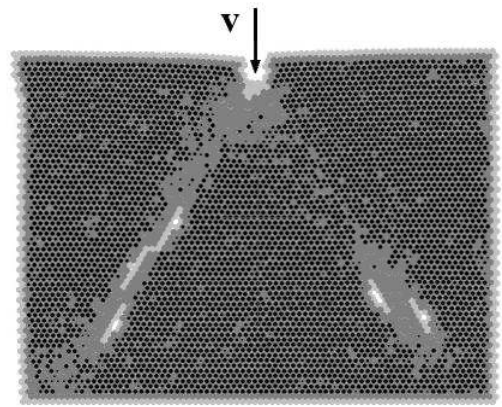}
\caption{\label{f2} Geometry of 2D computer experiment – positions of atoms at the time 0.21ns. More bright hue corresponds to higher potential energy of particles near dislocation cores.}
\end{figure}

At present, however, there are no experimental methods to registrate the motions of separate atoms. At the same time, the molecular dynamics methods allow to simultaneously simulate motions of all atoms of the lattice in detail. As an example let us consider the simulation of indentation into a 2D copper crystallite consisting of 80*70 atoms. The indenter (three atoms under arrow) moves with a constant velocity of 5m/s (see Fig. \ref{f2}).
 The lowest row of atoms is motionless. All lateral faces are free. The sample consists of copper atoms interacting through the Lennard-Jones potential. A detailed general formulation of the problem can be found in Ref. \onlinecite{m06}.

\begin{figure}
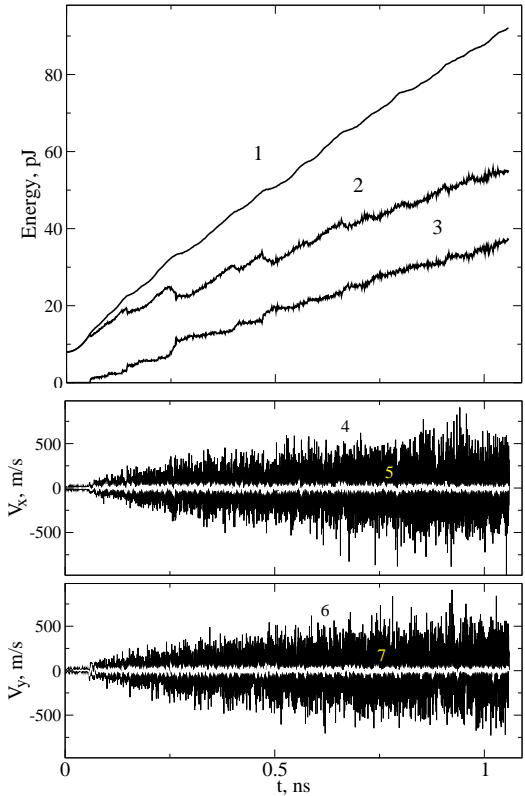

\hspace{0.06 cm}
\includegraphics [width=2.63 in] {fig_3a}
\vspace{0.09 cm}
\includegraphics [width=2.7 in] {fig_3b}
\caption{\label{f3} Time scanning: 1-3 – total internal, potential and kinetic energy, consequently; 4, 5 – X- and Y-components of velocity of an arbitrary particle in the centre of the model (light lines are low-frequency records of particle velocity after filtration)}
\end{figure}

At some moments the dislocations emerge in the monatomic specimen, as it is shown in fig. \ref{f2}.  This is pointed by bends of the curve 1 in Fig. \ref{f3} for time dependent total internal energy
 or by the jumps of total potential and kinetic energies (curve 2 and 3). The external work performed during the indentation increases total energies of all kinds. Further, for t = 0.055ns; 0.14ns; 0.24ns; 0.39ns and so on the potential energy decreases sharply. A part of this energy transforms into defect energy and the remaining part must dissipate into heat.

\begin{figure}
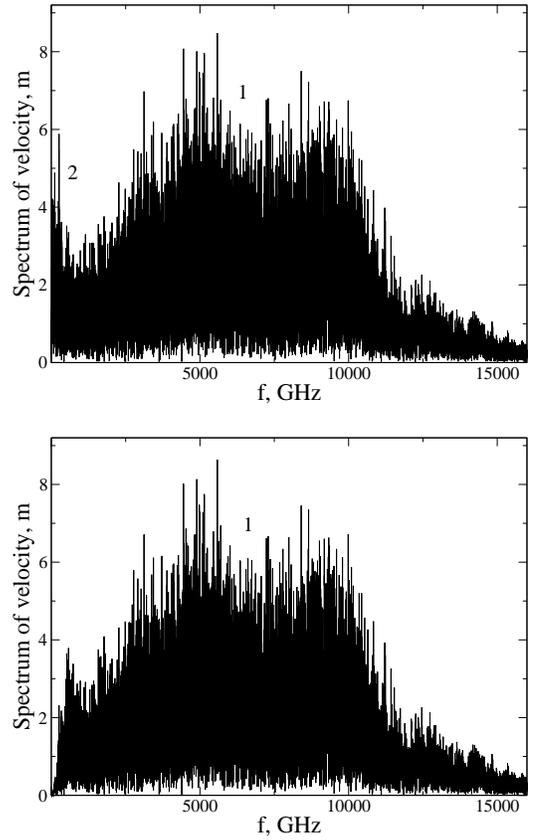

\includegraphics [width=2.7 in] {fig_4a}
\vspace{0.4 cm}
\includegraphics [width=2.7 in] {fig_4b}
\caption{\label{f4} Spectrum of complete record for velocity of a particle: 1 – main high-frequency region, 2 – low-frequency peak.}
\end{figure}

Bottom subplots in Fig. \ref{f3} present the time dependences of the X- and Y-components of velocity for an arbitrarily chosen particle of the system (near the center of the model), which actually are records of its thermal motion (incurves 4 and 6).
From a formal point of view, the thermal motion can be treated as a high-frequency signal which is close to a harmonic one for separate interatomic (bond) vibrations. The spectrum severely oscillates because of the interference of a large set of high-frequency phonons (see region 1 in Fig. \ref{f4}a).
 In the low-frequency region a large peak 2 stands up sharply against the rest of the spectrum. One can assume that its presence is connected with the regular low-frequency vibrations in the system.

Having done the low-frequency filtration with a running average method over 200 time steps one can separate the low-frequency vibration modes (bright curves 5 and 7 against main black signals). To make sure that these low-frequency vibrations are connected namely with the peak 2 of the spectrum of Fig. \ref{f4}a, the filtrated record was subtracted from the total record and for this difference signal the spectrum was calculated again (Fig. \ref{f4}b).
As seen, the low-frequency peak vanishes from the spectrum of the difference record.

Comparing curves 2, 3, 5 and 7 one can immediately see that the initiation of the low-frequency vibrational excitations coincides in time with the generation of dislocations and they can be in a sense associated with strongly nonlinear acoustic emission. On the other hand, they can be treated as nonequilibrium phonons with amplitudes far exceeding the amplitudes of corresponding low-frequency range of the spectrum which would be realized in thermal equilibrium (compare spectra of fig. \ref{f4}a and fig. \ref{f4}b). Thus, using the low-frequency filtration, it is possible to separate, in the pure state, the nonequilibrium subsystem (light curves 5 and 7) and the equilibrium subsystem (the difference signal) in a common thermal motion. The acoustic waves inelastically scatter by equilibrium high-frequency vibrations and, as a result, became damped. Decay of these waves is just a process of the nonequilibrium state relaxation. As a result, the energy of the acoustic waves gradually transforms into the energy of the equilibrium thermal motion leading to a slow increase of equilibrium temperature and entropy. 

\begin{figure}
\includegraphics [width=2.7 in] {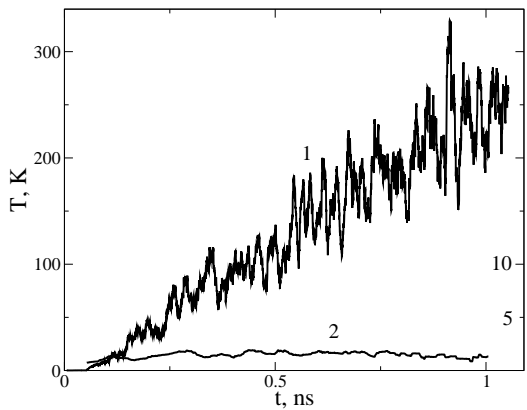}
\caption{\label{f5} Time scanning: 1 – temperature of equilibrium subsystem  ; 2 – temperature of non-equilibrium subsystem. For comparison, the scale of curve 2 (the right scale) is 10 times magnified.}
\end{figure}

During the indentation dislocations emerge repeatedly in time. That's why some fraction of the low-frequency nonequilibrium wave packets is always present in the integral thermal motion. On one hand, they are generated as a consequence of the mentioned processes of formation of defects, on the other hand, they are constantly going to the equilibrium subsystem. By averaging the square of particle velocity, that is the kinetic energy, for the difference record over time window containing 1000 time steps, we obtain an equilibrium temperature-like variable (curve 1, Fig. \ref{f5}).
Such averaging window contains nearly 20 periods of the high-frequency vibrations, and it seems to be enough to smooth of random fluctuations of the thermal motion. As seen, the equilibrium temperature increases as the energy of nonequilibrium low-frequency vibration transforms into the equilibrium high-frequency motion.

As the low-frequency acoustic wave differs from the high-frequency phonons only in the time scale then applying the same averaging procedure but over a larger time window containing in the given example 10000 time steps it is possible to definite an analogous variable for the non-equilibrium subsystem (curve 2, Fig. \ref{f5}).
 It can be given a sense of a nonequilibrium temperature or temperature of the nonequilibrium subsystem. Thus, the digital filtration allows us to separate equilibrium and non-equilibrium components in the initial data and to calculate their temperature parameters.

Now, equilibrium $S$ and nonequilibrium $\tilde{S}$ entropies can be defined as variables thermodynamically conjugate to the two kinds of temperatures described above. Production of nonequilibrium entropy will be described by kinetic equation of a form close to (\ref{a9}):
  \begin{equation}\label{a11}
\dfrac{\partial \tilde{S}}{\partial t}=-\gamma_{\tilde{S}} (T_{0} -\tilde{T_{st}}-T_{1}\tilde{S}).
  \end{equation}

Here $T_{0}$ is the temperature of internal sources at dislocation generation such that $\gamma_{\tilde{S}}T_{0}$ is the power of entropy sources, $\tilde{T_{st}}$ is the temperature of internal sinks of nonequilibrium heat dissipation such that $\gamma_{\tilde{S}}\tilde{T_{st}}$ is the power of entropy sinks. It also is a stationary temperature value, to which the system tends during its relaxation. The last term in Eq. (\ref{a11}) is a relaxation one, it describes transition of the entropy from non-equilibrium subsystem to equilibrium one because of the absorption of the low-frequency vibrations during their interaction with the equilibrium high-frequency vibrations.

\section{THERMODYNAMIC IDENTITY}

The results obtained in the previous section and their interpretation allow us to make some useful generalizations. In the thermodynamics and theory of self-organization, the subdivision of entropy change is well known \cite{gp71}:
  \begin{equation}\label{a12}
\bigtriangleup S=\bigtriangleup_{e} S+\bigtriangleup_{i} S,
  \end{equation}
where $\bigtriangleup_{e} S$ is the entropy change owing to heat flux from external sources (thermostats). The heat change is globally non-equilibrium but it is a locally equilibrium process. This component is positive at heating and negative at cooling. The component $\bigtriangleup_{i} S$ is the entropy production owing to irreversible phenomena from internal sources. Namely, this part of the entropy specifies the second law of thermodynamics and is always non-negative. The component is of a dual nature. On the one hand, this contribution to entropy is caused by non-equilibrium and irreversible nature of phenomena. On the other hand, this entropy part converts constantly to equilibrium form. Therefore, it is pertinent to present this part of the entropy as a sum:
  \begin{equation}\label{a13}
\bigtriangleup_{i} S=\bigtriangleup_{i}^{e} S+\bigtriangleup_{i}^{n} S,
  \end{equation}
where $\bigtriangleup_{i}^{e} S$ is the part of produced entropy succeeded in transforming to the equilibrium form during external action, $\bigtriangleup_{i}^{n} S$ is the part of produced entropy, which remains in the nonequilibrium form and should relax later. The former part $\bigtriangleup_{i}^{e} S$ became indistinguishable from another part of the equilibrium entropy (thermal degeneration) which had already been existing in the system before the action. Thus it makes sense to combine them denoting
  \begin{equation}\label{a14}
\bigtriangleup_{eq} S=\bigtriangleup_{e} S+\bigtriangleup_{i}^{e} S,
  \end{equation}
so that the total entropy change is:
  \begin{equation}\label{a15}
\bigtriangleup S=\bigtriangleup_{eq} S+\bigtriangleup_{i}^{n} S,
  \end{equation}

In such a form the entropy production has been firstly presented in this paper. Usually, one neglects the non-equilibrium part of the entropy $\bigtriangleup_{i}^{n} S$. Entropy production is accompanied by heat production, which in the first approximation can be presented as a linear combination of the entropy change of both kinds:
  \begin{equation}\label{a16}
\bigtriangleup Q=T\bigtriangleup_{eq} S+\tilde{T}\bigtriangleup_{i}^{n} S,
  \end{equation}
where $T$ and $\tilde{T}$ are some coefficients of proportionality with a dimension of temperature. Coefficient $T$ has the physical sense of the equilibrium temperature. Coefficient $\tilde{T}$ can be assigned a non-equilibrium temperature sense or a sense of the temperature of non-equilibrium subsystem. 

The total change of the internal energy under the external actions and owing to the change of internal structure as a consequence of relaxation processes can be written as a linear combination of changes of all independent variables:
  \begin{equation}\label{a17}
dU=V\sigma_{ij} d \varepsilon_{ij}^{e}+TdS+\tilde{T}\bigtriangleup \tilde{S}+\sum_{l=1}^{N} \varphi_{l} \bigtriangleup H_{l},
  \end{equation}
where more compact notation $dS=\bigtriangleup_{eq} S$ and $\bigtriangleup\tilde{S}=\bigtriangleup_{i}^{n} S$ is used. The first term is the change of the internal energy coinciding with the change of the elastic energy, $\sigma_{ij}$ is the stress tensor, $\varepsilon_{ij}^{e}$ is elastic deformation tensor (Coleman at all, 1967). The last term stands for the change of the internal energy owing to defect subsystems, $H_{l}$ is the number of $l$-type defects, $\varphi_{l}$ is the factor having the sense of defect energy. Here it is taken into account that the internal energy, the elastic deformation and the equilibrium entropy are functions of state, therefore their decrements are perfect differentials.

If, following the Landau's idea, one treats the non-equilibrium variables $\tilde{T}$ and $\varphi_{l}$ as generalized thermodynamic forces, for which relationships of the type (\ref{a3})
  \begin{equation}\label{a18}
\tilde{T}=\dfrac{\partial U}{\partial \tilde{S}},  
\quad \varphi_{l}=\dfrac{\partial U}{\partial H_{l}},
  \end{equation}
are true, then in expression (\ref{a17}), the increments of the independent variables $\bigtriangleup \tilde{S}$ and $\bigtriangleup H_{l}$  can be replaced by their exact differentials:
  \begin{equation}\label{a19}
dU=V\sigma_{ij} d \varepsilon_{ij}^{e}+TdS+\tilde{T}d \tilde{S}+\sum_{l=1}^{N} \varphi_{l} d H_{l},
  \end{equation}

Thus, the increment of the internal energy can be expressed in a perfect differential form and consequently, we can introduce an extended non-equilibrium state specified by a set of independent variables $\varepsilon_{ij}^{e}$, $S$, $\tilde{S}$ and $H_{l}$. A part of these variables $\varepsilon_{ij}^{e}$ and $S$ are equilibrated, the other part $\tilde{S}$ and $H_{l}$ are nonequilibrated. In this case, the internal energy is a function of these variables
  \begin{equation}\label{a20}
U=U(\varepsilon_{ij}^{^{e}}, S, \tilde{S}, H_{l})
  \end{equation}

To describe the isothermal phenomena, it is convenient to convert to the free energy $F=U-TS$, for which the differential form
  \begin{equation}\label{a24}
dF=V \sigma_{ij} d \varepsilon_{ij}^{e} - SdT +\tilde{T}d \tilde{S}+\sum_{l=1}^{N} \varphi_{l} d H_{l}
  \end{equation}
is true.

\section{DISCUSSION}

One can treat relationship (19) as a combination of the first and second laws of thermodynamics written in the form of identity. Previous attempts to formulate this law in the form of an identity are known \cite{e99,v01}. For example, in the Van's work the Gibbs relationship in the context of generalized thermodynamics with internal-state variables is presented in the form:
  \begin{equation}\label{a21}
du=Tds+\hat{t} \cdotp\cdotp d\hat{\varepsilon}-\vec{A} \cdotp d\vec{\alpha}
  \end{equation}
where $u$ and $s$ are densities of internal energy and entropy, $T$ is the absolute temperature, $\hat{\varepsilon}$ and  $\hat{t}$ are tensors of deformation (total) and stress, $\vec{A}$ is a vector thermodynamically conjugate to the vector of internal parameters $\vec{\alpha}$. The first two terms are useful attributes of the combined first and second laws of thermodynamics; the last term $\vec{A} \cdotp d\vec{\alpha}\geq 0$ quenches an excess energy. If one drops it, one obtains a common expression for the law in the form of inequality $du \leq Tds+\hat{t} \cdotp\cdotp d\hat{\varepsilon}$. In a case of the structureless body this inequality arises as a common part of the internal energy that has been taken in account twice, firstly in the form of irreversible work $\bigtriangleup A_{n}=\bigtriangleup A - \hat{t}\cdotp\cdotp d \hat{\varepsilon^{e}}$ and secondly in the form of internal heat released $T \bigtriangleup_{i}^{e} S + \tilde{T}\bigtriangleup_{i}^{n} S$. And this «redundant» energy is subtracted in the right-hand side of (\ref{a21}).  

However, as the total deformation $\hat{\varepsilon}$ is irreversible, the elastic stresses $\hat{t}$ can't be calculated as a derivative $\hat{t}=\partial u/\partial \hat{\varepsilon}$. Only the reversible (elastic) part of the deformation can suit the case $\hat{t}=\partial u/\partial \hat{\varepsilon^{e}}$ \cite{cg67}. Therefore, the structure of relationship (19) more correctly reflects the physical nature of the mutual energy transformations in internal processes. In relationship (19) two aspects of the law of conservation of energy are taken into account simultaneously. It is the law of conservation for a given reference volume interacting with the environment (the 1st law of thermodynamics) and the law of conservation or the law of transformation of energy in internal degrees of freedom $\bigtriangleup A_{n}=T \bigtriangleup_{i}^{e} S + \tilde{T}\bigtriangleup_{i}^{n}$. Moreover, in relationship (19) both structural aspect $\varphi_{l} d H_{l}$ and dynamic one $\tilde{T}d \tilde{S}$ are taken in account too.

The presentation of the combined first and second laws of thermodynamics for a specific case of unstructured solids and for isothermal processes in form of Jarzynski's identity \cite{j97a,j97b}:
  \begin{equation}\label{a22}
\exp(-\dfrac{\bigtriangleup F}{kT})=\langle \exp(-\dfrac{W}{kT}) \rangle
  \end{equation}
is also known. Here the angular brackets denote averaging over all states of the system, $\bigtriangleup F$ is the free energy increment in an isothermal process, $W$ is the work performed for the system. If the increment $\bigtriangleup F$ and the work $W$ are small then, expanding the exponents in a series and restricting ourselves to the first terms of the series, we obtain:
  \begin{equation}\label{a23}
dF=\langle W \rangle - \dfrac{\langle W^{2} \rangle}{2kT} = V \sigma_{ij} d \varepsilon_{ij}-\dfrac{\langle W^{2} \rangle}{2kT}= V \sigma_{ij} d \varepsilon_{ij}^{e}.
  \end{equation}

The first term $\langle W \rangle$ presents the total work performed for the system; the second term $\langle W^{2} \rangle / 2kT$ presents energy dissipated as a result of irreversible internal processes, consequently their difference is the reversible elastic energy. 

For an isothermal process $dT=0$ with neglecting both transitional thermal processes $d \tilde{S}=0$ and contribution from structure of a solid $d H_{l}=0$, we obtain from (\ref{a24}) $dF=V \sigma_{ij} d \varepsilon_{ij}^{e}$ that coincides with the Jarzynski's identity in the limit (\ref{a23}).

\section{OUTLINE FOR APPLICATION TO SEVERE PLASTIC DEFORMATION}

It is of interest to consider how the obtained above general relationships can be applied to solving practical problems. One of such problems is a change of the internal structure of metals processed by severe plastic deformation (SPD) \cite{via00,kc06,zlz02}. During SPD the grain boundaries are intensively produced resulting in refinement and fragmentation of the grains. The refinement of grains at the expense of multiplication of their boundaries is the main process of SPD and the grain boundaries are the main structural defect \cite{zlz02}. At the same time, the main building «material» for the grain boundaries are dislocations and dislocation piles-up. That's why, for the description of SPD these two kinds of defects or the two levels of defectiveness need to be taken in account as a minimum. With quadratic terms remained in the expansion of the internal energy
  \begin{equation}\label{a25}
U=U_{0}+\varphi_{0}H-\dfrac{1}{2}\varphi_{1}H^{2}+\tilde{\varphi_{0}} \tilde{H}-\dfrac{1}{2} \tilde{\varphi}_{1} \tilde{H}^{2}+\phi H \tilde{H},
  \end{equation}
we write down the corresponding kinetic equations (\ref{a9}) for this case:
\begin{eqnarray}\label{a26}
\nonumber
\tau_{H} \dfrac{\partial H}{\partial t}=\varphi_{0}-\varphi_{1} H + \phi \tilde{H},  \\
\tau_{\tilde{H}} \dfrac{\partial \tilde{H}}{\partial t}=\tilde{\varphi}_{0}-\tilde{\varphi}_{1} \tilde{H} + \phi H.
\end{eqnarray}

Here variables relating to the dislocations are marked by tilde. The last terms describe the mutual influence of different levels. In an adiabatic limit $\tau_{\tilde{H}}\ll\tau_{H}$ the equilibrium value of dislocation number (or density) is derived from stationary condition for the second Eq. (\ref{a11}):
  \begin{equation}\label{a27}
\tilde{H}=\dfrac{1}{\tilde{\varphi}_{1}}(\tilde{\varphi}_{0}-\phi H).
  \end{equation}
This stationary value can be used to exclude the dislocation number from the first Eq. (\ref{a26}):
  \begin{equation}\label{a28}
\tau_{H} \dfrac{\partial H}{\partial t}=\varphi_{0}-\varphi_{1} H +\dfrac{\phi}{\tilde{\varphi}_{1}}(\tilde{\varphi}_{0}-\phi H)=\varphi^{ef}_{0}-\varphi^{ef}_{1} H.
  \end{equation}

Thus, the account of dislocations in the adiabatic approach has brought only the renormalization of theory constants. Let us consider the next level of approximation, we suppose, namely, that the right-hand part of the second Eq. (\ref{a26}) is not strictly zero, but equals a small constant ε, which depends parametrically on $H$ (quasi-adiabatic approximation):
  \begin{equation}\label{a29}
\tilde{\varphi}_{0}-\tilde{\varphi}_{1} \tilde{H} + \phi H=\varepsilon
  \end{equation}
Integrating the second Eq. (\ref{a26}) we obtain
  \begin{equation}\label{a30}
\tilde{H} =\dfrac{1}{\tau_{\tilde{H}}} \varepsilon t + \dfrac{1}{\tilde{\varphi}_{1}}(\tilde{\varphi}_{0}+\phi H).
  \end{equation}
Substituting this relationship into the first Eq. (\ref{a26}) and taking $\varepsilon=\varepsilon_{0} + \varepsilon_{1} H$ we find
  \begin{equation}\label{a31}
\tau_{H} \dfrac{\partial H}{\partial t}=\varphi^{ef}_{0}-\varphi^{ef}_{1} H + \dfrac{\phi}{\tau_{\tilde{H}}}\varepsilon_{0} t-\dfrac{\phi}{\tau_{\tilde{H}}}\varepsilon_{1} H t.
  \end{equation}
Supposing formally $h=\nu/k$ and $t=e/V$, and also $\varphi^{ef}_{0}=-\kappa \tau_{H} e_{0} V/k$, $\varphi^{ef}_{1}=-\kappa \tau_{H} e_{0} V$, $\varepsilon_{0}=-\kappa \tau_{H} \tau_{\tilde{H}} V^{2}/k\phi$, $\varepsilon_{1}=-\kappa \tau_{H} \tau_{\tilde{H}} V^{2}/\phi$ we get an equation close to the Rybin's kinetic equation \cite{r91}:
  \begin{equation}\label{a32}
\dfrac{1}{\kappa}\dfrac{d\nu}{de}=-e_{0}+e_{0}\nu+e-\nu e.
  \end{equation}

Note that the first two terms in Eqs. (\ref{a31}) and (\ref{a32}) have opposite signs. The first term in (\ref{a32}) can be treated as annihilation of grain boundaries, going «by itself» at a constant speed. The second term describes the generation of boundaries, the speed of generation being the higher the more boundaries (or fragments) are by this time present in a solid. For $e_{0}\approx0.2$ \cite{r91}, it can be argued that the contribution of the first two terms to kinetics of fragmentation is small. Within this accuracy Eqs. (\ref{a31}) and (\ref{a32}) coincide.

Thus, in the framework of the approach, the well-known and well-approbated kinetic equation for the description of the grain fragmentation during SPD has been derived. At the same time, the approach has a more general context that permits to take the influence of thermal processes into account and to develop two-mode approximations with the expansion of thermodynamic potentials into a  higher power series in terms of defectiveness of solids \cite{m09}.

\section{CONCLUDING REMARKS}

In the paper, a phenomenological approach based on generalization of Landau technique to cases of quick-passing nonequilibrium phenomena, for example, severe plastic deformation, is considered. For quick-passing processes thermal fluctuations have no time to exert essential influence and it is possible to consider the problem in the mean-field approximation. In the centre of the approach there is not an abstract order parameter but a structural defect with its real parameters – the quantity (density) of defects and the average energy of a defect. The analysis from these positions on the example of a version of the theory of vacancies shows that the known Landau-Khalatnikov kinetic equation can be used in this case only if the dependence of vacancy energy on density of vacancies is neglected. In the case of an arbitrary dependence of vacancy energy on density of vacancies it is necessary to use a more general form of kinetic equations symmetric with respect to using the internal and free energies. In this case, the density of defects and defect energy are related by a symmetric differential dependence of type (\ref{a8}) and (\ref{a9}). As the defect energy in the steady state is not equal to zero then the extreme principle of equality to zero of the derivative of free energy  with respect to order parameter (in our case the number or the density of defects) breaks down. This principle need be substituted with principle of the tendency to a steady state. Steady-state characteristics can not be determined in the framework of phenomenological approach, here statistical and microscopic approaches are required.

A form of kinetic equations can be generalized to all types of regularly or randomly distributed defects and to all types of nonequilibrium parameters including thermal variables. On an example of computer experiment it has been shown that in the case of quick-passing processes the irreversibility is not so much connected with fluctuation phenomena, as with dynamical ones of acoustic emission type at the formation and motion of defects. In this case, the acoustic emission over all structural levels is a part of the relaxation process \cite{ptz07,i04,k06} that permits to describe the entropy generation (\ref{a11}) in the same form as generation of defects (\ref{a9}).

The obtained results are generalized in the form of a thermodynamic identity which properly combines the 1st and the 2nd laws of thermodynamics in a finite-difference form. The known Jarzynski thermodynamic identity for isothermal processes expressed in terms of free energy coincides, in going to finite differences, with the proposed identity.

To illustrate the applicability of the approach a problem of description of nonequilibrium phenomena occurring under severe plastic deformation is considered. In the framework of phenomenological one-mode (approximation quadratic with respect to energy) model taking into account the two levels of defect structure – dislocations and grain boundaries, the known for this case kinetic Rybin equation is derived \cite{r91}. The approach permits to consider more complex dependences of thermodynamic potentials on basic parameters of the model. For example, in the framework of two-mode approximation using the polynomials of the fourth power for presentation of thermodynamic potentials the severe plastic deformation can be represented as a structural phase transition of the 1st type \cite{m09}.

\begin{acknowledgments}
The work was supported by the budget topic 0106U006931 of NAS of Ukraine and partially by the Ukrainian state fund of fundamental researches (grants F28.7/060). The author thanks A. Filippov for helpful discussions.
\end{acknowledgments}


\end{document}